\documentstyle[12pt,aaspp4,psfig]{article}
\begin{document}

\title{Time Dilation of BATSE Gamma-ray Bursts}
\author{Theodore T. Lee\altaffilmark{1} and Vah\'{e} Petrosian\altaffilmark{2}}
\affil{Center for Space Science and Astrophysics,
Stanford University, Stanford, CA 94305-4060}

\altaffiltext{1}{Department of Applied Physics}
\altaffiltext{2}{Departments of Physics and Applied Physics}

\begin{abstract}
We perform several nonparametric correlation tests on the BATSE 3B
data to search for evidence of cosmological time dilation.
These tests account for the effects of data truncation due to
threshold effects in both limiting brightness and limiting duration,
enabling us to utilize a larger number of bursts
than in previous analyses.  We find little significant evidence for
correlation between various measures of peak intensity and duration,
but the tests cannot conclusively rule out time dilation factors of
2 or less without more data.  There is stronger 
evidence for a positive correlation between fluence and duration,
which if confirmed would rule out simple no-evolution cosmological models
unless there is a strong intrinsic correlation between the total
radiant energy and the duration of bursts.
\end{abstract}

\section{Introduction}

Observations taken by the BATSE instrument aboard the {\em Compton
Gamma-Ray Observatory} have shown that the angular distribution of
gamma-ray bursts (GRBs) is isotropic (Meegan et al. 1992), while the
$\log{N}$-$\log{S}$ distribution flattens for weaker bursts.  Such a
distribution would be expected if bursts were of cosmological origin.
Paczy\'nski (1992) and Piran (1992) suggested that if bursts originated
at cosmological distances, their light curves should be stretched due
to cosmological time dilation.
If bursts were standard candles, dimmer (hence more distant)
bursts would be time-dilated more than brighter (hence less distant)
bursts, by a dilation factor 
$(1+z_{dim})/(1+z_{bright})$, where $z_{dim}$ and $z_{bright}$ are
the redshifts.

However, the expected redshift range of order unity would result in
a time dilation factor of a few while the 
burst durations cover a large dynamic range from tens of
milliseconds to hundreds of seconds (Kouveliotou et al. 1993). 
Therefore a time dilation effect can only be detected
statistically.  Norris et al. (1994, 1995) 
(hereafter N1 and N2) searched for time dilation
effects by dividing the bursts into groups based on their peak
count rate $C_P$ and comparing some measure of burst duration with peak
count rate.  They found evidence that brighter bursts had shorter
durations than dimmer ones, and that the difference between the average
durations of bright and dim bursts was consistent with a time dilation
factor of about 2.  This factor of 2 time dilation is what would be
expected 
for simple cosmological models if the peak luminosity of the
bursts were independent of their duration or redshift. This effect
would be easier to detect if the peak luminosity function of the bursts
was narrow; in other words, if the burst peak luminosity was a 
``standard candle.''  Fits of the $\log{N}$-$\log{C_P}$ distributions
to cosmological models with the standard candle peak luminosity
assumption
(Piran 1992; Mao \& Paczy\'nski 1992; Wickramasinghe et al. 1993;
Fenimore et al.
1995; Azzam \& Petrosian 1996) obtain a mean
redshift $z_{dim} \sim 1$ so that the dilation factor would be about 2
($z_{bright} \ll 1$).

However, Band (1994) has cautioned that an intrinsic burst luminosity
function could easily produce similar effects.  Yi \& Mao (1994) also
noted that relativistic beaming in either Galactic halo or cosmological
models can produce flux-duration relationships
that might be consistent with the reported effects.
Wijers \& Paczy\'nski
(1994) suggested a way to distinguish between anticorrelations between
flux and duration produced by cosmological time dilation and those
produced by a decrease in burst density with distance, which is
needed in a local extended halo model if the shape of the luminosity
function is independent of distance.  
They cautiously concluded that the data from the first
BATSE catalog is more consistent with a cosmological interpretation.
On the other hand, as shown by Fenimore (1996), the agreement of the
time dilation results with the
cosmological models is destroyed by the fact that the burst duration
$T$ (or the pulse widths) appears to be well correlated with photon
energy $E$; $T \propto E^{-\alpha}$ (Fenimore et al. 1995).  Since higher redshifts correspond
to higher energies of emitted photons, the expected time dilation effect
will not yield a $T\propto (1 + z)$ relation but rather
$T\propto (1 + z)^{1-\alpha}$.  For $\alpha = 1/2$ this would require higher redshifts 
($z_{bright} \approx 1$ and $z_{dim} \approx 6$),
destroying the good agreement with
the $\log{N}$-$\log{S}$ results.  Furthermore, there is even  now disagreement
about the reality of the observed time dilation.
For example, Mitrofanov et al. (1994; 1996) compared averaged time histories of
weak and strong bursts detected by BATSE and found no evidence for any
time dilation.  Similarly, Fenimore et al. (1995) noted that a 
different method of analysis, which agrees with the results of
N1 and N2 when used for the same set of data, gives a much smaller
time dilation factor (1.3 instead of 2) for a larger set of data.
Recently, Norris et al. (1996) has reported that the larger data
set is consistent with observed time dilation factors of 1.5--1.7.

It is clear that despite the numerous works published on the subject,
time dilation of gamma-ray bursts remains controversial.  Here we present
new results on this topic which differ from previous works in
two important ways.  Previous
studies of burst time dilation have been limited to bright, long
duration bursts.  This selection avoids the biases against detection by
BATSE of bursts with durations shorter than the trigger integration
time, and against weaker bursts due to the variability of the 
threshold photon count rate.  As a result,
these studies utilize only a small percentage ($\sim 20\%$) of the total
number of bursts.  
In this work we
attempt to extend the time dilation analyses to a
sample of bursts which is larger by about a factor of two,
by properly accounting for the effects of
variable thresholds and the short duration bias.  
These biases and the methods to account for them are described in
detail in several of our previous publications (Efron \& Petrosian 1992;
Petrosian 1993; Lee, Petrosian, \& McTiernan 1993; Petrosian, Lee, \& Azzam 1994).
For a complete review see Lee \& Petrosian (1996; hereafter LP).

Second, all of the previous studies use the peak photon count rate
or the peak flux as a measure of the distance.  
This usage assumes that the peak photon luminosity
is independent of the burst duration or distance. In addition, a
narrow distribution of the peak luminosity is required for the
detection of the small time dilation effect.  In other words the
peak photon luminosity should be nearly a standard candle
for the previous studies to be valid.  This
seems not very likely considering the large dispersion in
the duration and pulse shapes of GRBs.  We believe that it is more
likely that the total energy (or total number of photons) emitted
by a burst is a standard candle, so that the energy or photon
fluence will be a better measure of the distance to the bursts.
Because bursts are generally detected on the basis of their peak flux,
it has been difficult in the past to examine unbiased fluence 
distributions.  However, as shown by Petrosian \& Lee (1996; hereafter PL)
the methods mentioned above work equally
well with fluence, so
we carry out our tests using both peak fluxes and fluences.  Furthermore,
because of the complex and varied burst pulse shapes, it is not
clear what measure of the time structure, or which of the
several available time scales associated with the pulse profiles,
would be a reliable measure of redshift in a cosmological scenario.  
In addition to using the durations $T_{50}$ and $T_{90}$ provided
in the BATSE catalog, we also use an effective duration defined by
the ratio of the total energy released to the peak luminosity,
which we believe to be a robust measure of duration and therefore
redshift.  A brief summary of our method is given in \S \ref{sec-analysis},
the choice of test variables is discussed in \S \ref{sec-choice},
the test results are presented in \S \ref{sec-testresults}, 
and a discussion and summary of our conclusions are given in \S \ref{sec-dilconclusion}.

\section{Analysis Method} \label{sec-analysis}

The problem of searching for time dilation effects can be thought of as a
search for a correlation between two variables, one of which is
some measure of burst duration while the other is some measure of
burst brightness.  Due to the detection biases against short duration
and weak bursts,
the practice in previous dilation studies has been to
examine relatively bright, long duration bursts.

However, it is possible to extend such a test to a much larger
sample if the observational selection criteria or data truncations
are well defined, because there exist methods to test for 
correlations in the presence of such truncations.
A simple test which is easily applied
to burst data is the $t_w$ test described by Efron \& Petrosian (1992).
Briefly, the test relies on the concept of the {\em associated set} of
points for each data point.  For an untruncated data set, the associated
set of data points is the entire set of data points.  For a truncated
data set, the associated set differs for each data point and is
defined as the largest subset of data points for which there is
no truncation.  Each data point can then be assigned a rank amongst
the points comprising its associated set.  If a correlation exists,
then the ranks will be correlated.  Because the expected distribution of ranks
for an uncorrelated data set is well-defined no matter what the
individual distribution functions of the variables are, the test is
completely nonparametric.  Efron \& Petrosian's $t_w$ test
has the following properties:
\begin{enumerate}
\item The scalar statistic $t_w$ calculated from the data could be positive or
negative, with the
sign indicating correlation (+) or anticorrelation (-).
\item The value of $t_w$ gives the probability 
$P(t_w) = $erfc$(|t_w|/\sqrt{2})$ that the data were drawn from an 
uncorrelated distribution.  
In other words, $t_w$ is distributed normally and can be
interpreted as the number of standard deviations away from 
the expected result of the test for perfectly uncorrelated data.
In this sense $t_w$ can be used as an error estimate.
\item Each data point is weighted, and the weights can be chosen
such that the $t_w$ test becomes equivalent to standard tests
such as Kendall's $\tau$ for simple truncations.  Throughout
this paper we give every data point the same weight, which we
indicate notationally by referring to this variation of the $t_w$ test 
as $t_1$.
\end{enumerate}
Thus, to test the time dilation hypothesis, it becomes merely a matter
of choosing the variables to test and defining the data truncations.
As discussed below, these tasks can be far from straightforward.

\section{Choice of Variables} \label{sec-choice}

The BATSE catalog provides several observed quantities which could be
used as a measure of distance to the burst.
It is not immediately obvious which measures are the most suitable, so we
consider them carefully below.

\subsection{Peak Photon Flux}

The most observationally tractable
quantities are the average peak photon count rate $\bar{C}_P = C_{max}/\Delta t$, where the
average is over the trigger time $\Delta t$.
The burst selection criterion is that $\bar{C}_P > \bar{C}_{lim}$,
where $\bar{C}_{lim}=C_{min}/\Delta t$
is the threshold value set 
by the background count rate.  The BATSE catalog gives the values of 
$C_{max}$ and $C_{min}$ for three trigger times $\Delta t=64,256$, and 1024~ms.
The rate $\bar{C}_P$ is related to
the peak photon luminosity through the instrument response and
its dependence on the angular location of the burst.  A better
measure is the average peak photon flux $\bar{f}_P$, which is directly
related to the average peak photon luminosity $\bar{F}_P$ as
$\bar{f}_P = \bar{F}_P/4\pi d_L^2$, where $d_L$ is the appropriate
luminosity distance.  In the cosmological scenarios $d_L$ 
depends on the redshift and the model parameters.  
Thus, in the rest frame of any particular burst the
time interval over which the peak flux is averaged depends on its redshift.
As described in LP the threshold flux for each
burst can be obtained from the catalog as 
\begin{equation}
\bar{f}_{lim} = \bar{f}_P(\bar{C}_{min}/\bar{C}_P) = \bar{f}_P({C}_{min}/{C}_{max}). \label{eq:flimtrue}
\end{equation}
These time-averaged flux measures are subject to the 
short duration bias for bursts with durations less than the
trigger interval $\Delta t$.  It was shown in LP
(see also Petrosian, Lee, \& Azzam 1994) that 
an approximate
way of correcting for this bias is to define our best estimate of
the ``true'' or instantaneous peak flux as
\begin{equation}
\hat{f}_P = \bar{f}_P(1+\Delta t/T), \label{eq:fptrue}
\end{equation}
where $T$ is the ``true'' duration of the burst.  Typically some
measure of the duration is used to estimate this true duration, so
that equation (\ref{eq:fptrue}) becomes
\begin{equation}
\hat{f}_P(T_x) = \bar{f}_P(1+\eta \Delta t/T_x), \label{eq:fptruex}
\end{equation}
where $T_x$ is some observationally tractable duration measure and 
$\eta \equiv T_x/T$.   LP found that $\eta \approx 0.5$ fit the
BATSE data for $T_x=T_{50}$ in a statistical sense. 
A similar correction can be used to define an estimate of the instantaneous
peak 
photon count rate $\hat{C}_P$ from the observed average value $\bar{C}_P$.  
Note that these transformations remove not only the duration bias
but also the ambiguity of the redshift dependence of the
rest frame trigger interval.  In the equation $\hat{f}_P = F_P/4\pi d_L^2$,
the only redshift dependence is through $d_L$.

\subsection{Fluence} \label{sec-fluencesection}

As mentioned before it seems unlikely that $F_P$ is a standard candle.
Another possible candidate is the total radiated energy $\varepsilon$ or
the total number of emitted photons.  Unlike $F_P$, neither of these quantities would be
affected by the boosting due to the large bulk Lorentz
factors that would be appropriate for a cosmological fireball scenario
(Paczy\'nski 1986; Goodman 1986; Meszaros \& Rees 1993, 1994).  
The appropriate observational measure found in the
BATSE catalog is the energy fluence ${\cal F} = \varepsilon/4\pi \tilde{d}_L^2$,
where $\tilde{d}_L = d_L/\sqrt{1+z} = d_m(1+z)$ where $d_m$ is the
metric distance.  As described in PL
the threshold on the fluence is obtained as 
\begin{equation}
{\cal F}_{lim} = {\cal F}({C}_{min}/{C}_{max}).
\end{equation}
Like $\hat{f}_P$ and $C_P$, the fluence ${\cal F}$ is also subject to a bias.
The bias now is against the detection of weak and long bursts
which have $\bar{C}_P$ to low
to exceed the threshold $\bar{C}_{lim}$.  Following the same
arguments used in LP which led to the correction of peak 
fluxes according to equation (\ref{eq:fptrue}), it can be
shown that 
\begin{equation}
{\cal F}_{lim} \approx \langle h\nu \rangle \bar{f}_{lim} (\Delta t + T),
\end{equation}
where $\langle h\nu \rangle$ is the average energy per photon in
the 50--300 keV range.  Because of the strong dependence of ${\cal F}_{lim}$
on $T$, we argue later in \S \ref{sec-flujust}
that a test of correlation between ${\cal F}$
and ${\cal F}_{lim}$ can be used to effectively test the duration--fluence
correlation.

\subsection{Observational Measures of Duration}

\subsubsection{Effective Duration} \label{sec-effdur}

As a measure of redshift we would ideally like to use the ``true'' 
duration $T$ of the burst, which is an observationally ill-defined
quantity.
There are several available observational measures $T_x$
of burst duration which could potentially be
used as measures of redshift.  
Two of these, $T_{90}$ and $T_{50}$, are listed in the catalog and
represent the time interval between the instances when the burst reaches
5\% and 95\% of its total fluence for $T_{90}$, and 25\% to 75\% of the fluence for
$T_{50}$.  As an alternative measure of duration we define an ``effective duration''
\begin{equation}
T_{eff}(T_x) = \frac{{\cal F}}{\hat{f}_P \langle h\nu \rangle} 
	= \frac{{\cal F}}{\bar{f}_P(1+\eta \Delta t/T_x)\langle h\nu \rangle}.
	\label{eq:teff}
\end{equation}
We calculate $\langle h\nu \rangle$ by assuming a power law energy
spectrum and using the ratio of 100--300 keV to
50--100 keV fluence to solve for the power law index.  This approximation
should not affect the correlation analysis described below unless
there is a very strong correlation between the spectral index and
fluence or flux.  In any case, the value of $\langle h\nu \rangle$ is
insensitive to the spectrum, varying by less than 10\% for simple 
power laws and by less than 25\% for Band (1993) type spectra.
We note, however, that there does exist a correlation between
spectral hardness and duration,
in that shorter bursts tend to have harder spectra (Kouveliotou et al.
1993).  If bursts were cosmological, their spectra should be
more redshifted the more distant they are.  Since a ``typical''
GRB tends to have a spectral steepening at about 150~keV (Band et al.
1993), a redshifted burst will be expected to have a softer spectrum,
and hence a smaller $\langle h\nu \rangle$ than a local burst.
Not correcting for this effect can only accentuate the time
dilation effect on the effective duration, because the smaller
values of $\langle h\nu \rangle$ for bursts at high redshift will
tend to make $T_{eff}$ longer due to the form of equation
(\ref{eq:teff}).  As will be seen below, we do not believe that
the results of the correlation tests warrant a correction for the
effect.

\subsubsection{Signal-to-Noise Bias}

Norris (1996) has shown that there are problems with using
$T_{50}$ and $T_{90}$ as estimators of the time dilation factors.
Consider bursts with light curves consisting of a
dominant spike and one or more smaller spikes separated by long
quiescent periods (for example, see the light curves of bursts 143,
219, 841, 1145, 1440 in the BATSE catalog).  Such bursts will be
assigned long values of $T_{90}$ due to the presence of the small
spike(s).  However, for bursts of similar time profiles but weaker
intensities these small spike features could be lost in the background
noise and would be considered as single spike bursts, leading to much shorter
values of $T_{90}$.  This signal-to-noise bias
would lead to a {\em correlation} between
$C_P$ and $T_{90}$, and would tend to cancel out a time dilation
effect.  To a lesser extent, $T_{50}$ suffers from this bias as well.
 We believe $T_{eff}$ may be a more stable 
measure of duration with respect to fluctuations in the background
noise level.  For instance, for multipeaked bursts such as those mentioned
above both $T_{50}$ and $T_{90}$ could jump discontinuously as a
function of signal-to-noise ratio, while $T_{eff}$ would change
gradually and often insignificantly.  On the other hand,
$T_{eff}$ does suffer from the necessity of estimating
the peak flux from time-averaged data, requiring some independent measure
of duration (e.g. using Eq. [\ref{eq:fptrue}] with
$2T_{50}$ in place of $T$).  However, this is a second-order effect
because for the majority of bursts $T_x \gg \eta \Delta t$ and hence
$T_{eff}$ is independent of $T_x$.

Figure \ref{fig:tcomp64} shows a comparison of the various definitions
of duration obtained with BATSE catalog data.  In these plots, we
show ratios of the various measures of duration as functions of 
$\bar{C}_P/\bar{C}_{lim}$, which should serve as an estimate of the
signal-to-noise ratio.  In the top two panels of the Figure, it can be seen from
the graphs that there is a systematic trend for bursts with smaller
signal-to-noise ratio to have smaller ratios of $T_{90}/T_{eff}$ and
$T_{90}/T_{50}$,
clearly demonstrating the signal-to-noise bias.  This effect can approach
a factor of 2, which is of the same order as the expected time dilation effect.
As shown in the bottom panel, $T_{50}$ is much less 
susceptible to the bias than $T_{90}$.  We have repeated this analysis
for a subset of bursts kindly provided to us by J. Norris, 
for which the $T_{50}$ and $T_{90}$ durations have been calculated only
after normalizing the peak intensities in order to remove the
signal-to-noise bias.  We denote these corrected duration estimates as 
$\hat{T}_{50}$ and $\hat{T}_{90}$.
Substituting these peak-normalized durations for $T_{eff}$ in the ratios
depicted in the Figure, 
we find trends similar to those shown in Figure \ref{fig:tcomp64}.
Comparing $T_{eff}$ to the peak-normalized measures of duration 
seems to show very little evidence for this trend, in agreement
with our conclusion above that $T_{eff}$ should not suffer as much
from the bias.  

Clearly, it would be preferable to use peak-normalized
burst duration estimates $\hat{T}_{90}$ and $\hat{T}_{50}$
instead of the BATSE values of $T_{90}$
and $T_{50}$ in our $t_w$ test.  However, the
very utility of our test lies in its ability to extend the
burst sample to include very weak bursts.  It is precisely
these bursts for which the peak-normalization procedure becomes
problematic.  
Lacking $\hat{T}_{90}$ and $\hat{T}_{50}$ for the larger sample of bursts
we wish to test, we take $T_{eff}$ to be the most robust available
duration measure.  We also note that in LP it was shown that corrections
for the short duration bias based on equation (\ref{eq:fptruex})
provided the best agreement between $\Delta t=64$~ms and $\Delta t=1024$~ms
duration distributions when $T_{eff}$ was used as an estimator for
$T_x$.

\section{Correlation Test Results} \label{sec-testresults}

We have performed the $t_w$ test for stochastic independence 
on several combinations of variables.  In all these tests,
we choose the standard weight vector $\vec{w} = (1,1,\ldots,1)$, so that
each data point is given the same relative importance (see Efron \& Petrosian
1992 for a discussion of weights in the $t_w$ test).
As a measure of distance we use the average peak flux $\bar{f}_P$,
the estimated true peak flux $\hat{f}_P$, and the fluence ${\cal F}$.
We do not consider $\bar{C}_P$ and $\hat{C}_P$, which give results
similar to $\bar{f}_P$ and $\hat{f}_P$, respectively.  As a measure of
redshift we use the durations $T_{90}$, $T_{50}$,
and $T_{eff}(T_{50})$ as defined above.
We carry out these tests for data obtained at all three
trigger times $\Delta t = 64$, 256, and 1024~ms separately.

\subsection{Direct Test for Peak Flux vs. Duration} \label{sec-peakvsdur}

A direct test of the correlation between the average peak flux
$\bar{f}_P$ and duration 
is complicated because the threshold $\bar{f}_{lim}$ varies and
a clear truncation boundary cannot
be delineated in the $\bar{f}_P$--$T$ plane.  Analysis of the
three-dimensional distribution involving $\bar{f}_P$, $\bar{f}_{lim}$,
and $T$ is required to properly take into account the variation of
$\bar{f}_{lim}$.  This point is discussed in  \S \ref{sec-indirect},
but 
for the sake of simplicity and clarity we first limit ourselves
to the two-dimensional case by selecting a subsample
of data which could be described with a single constant
threshold $\bar{f}_{lim}$.  An obvious choice is to limit the
subsample to the sources with $\bar{f}_P$ greater than or 
equal to the maximum value of the observed values of $\bar{f}_{lim}$.
In this case, all sources with $\bar{f}_P \leq \bar{f}_{lim,max}$
are excluded from the analysis.
A slightly better choice 
is to find the value of $\bar{f}_{lim} = 
\bar{f}_{lim,0}$ such that the truncation is kept simple 
and the number of data points is maximized.
In practice this amounts to limiting the data points to those with
$\bar{f}_{lim} \leq \bar{f}_{lim,0}$
and $\bar{f}_P \geq
\bar{f}_{lim,0}$.  
With a constant $\bar{f}_{lim}$,
the truncation boundaries in the $\bar{f}_P$--$T$ plane become parallel to
the axes and the problem reduces to one of simple truncation.
The $t_1$ test then reduces down to a simple rank order 
correlation test.  

\subsubsection{Average Peak Flux---Duration Correlation}

Using this truncation, we calculate $t_1($data) values to test
the correlation between the average peak flux $\bar{f}_P$ and various observational
measures of duration.
The first three rows of Table \ref{table:fpbar} show the values of
$t_1($data) for each of the three trigger times $\Delta t = 64,256,$
and 1024~ms, respectively.  The first obvious feature in these 
numbers is that the values of $t_1($data) are significantly
and consistently larger for correlations involving $T_{90}$
than for the other measures of duration, which give nearly
identical results.  The most likely explanation of this result
is that the $T_{90}$ values are underestimated at low values of
$\bar{f}_P$ due to the signal-to-noise bias, 
giving rise to a larger positive value for $t_1($data)
and an apparent correlation.  This result is in agreement with
the relations discussed in connection with Figure \ref{fig:tcomp64}
and the findings by Norris (1996).

\subsubsection{Corrected Peak Flux---Duration Correlation}

The second
feature of these $t_1($data) values is that they are larger for larger
values of $\Delta t$.  This result is most likely due to the
short duration bias mentioned in connection with equation (\ref{eq:fptrue}).
We may use the approximation
of equation (\ref{eq:fptrue}) to correct the average peak flux,
using $(100/x)T_x$ as estimates for the true duration $T$.
The magnitude of this correction increases with the ratio of
$\Delta t/T$, and is therefore largest for $\Delta t=1024$~ms at
$T \lesssim 1$~s.  The transformed data in the $\hat{f}_P$--$T_x$ plane is
no longer truncated by a single average peak flux limit.  Instead
the truncation is defined by 
\begin{equation}
\hat{f}_P > \hat{f}_{lim,0} = \bar{f}_{lim,0} (1 + \Delta t/T), \label{eq:dilfpflim}
\end{equation}
which cannot be described as a simple truncation (see Fig. 6 of LP).
With proper account for the truncation given by equation (\ref{eq:dilfpflim}),
the values of $t_1($data) obtained for these data are given
in the second three rows of Table \ref{table:fpbar}.
The values of $t_1($data) are now lower, especially for the
$\Delta t$=1024~ms data.  This result apparently confirms the
assertion that the differences between the data sets
corresponding to the three different values of $\Delta t$
are due to the short duration bias.  In particular the good
agreement found for $T_{eff}$ and the three values of $\Delta t$
indicates again that equations (\ref{eq:fptrue}), (\ref{eq:fptruex}),
and (\ref{eq:dilfpflim}) are fairly reliable.  The least biased of
these tests should be the one involving $T_{eff}(T_{50})$ and 
$\hat{f}_P(T_{eff})$ with $\Delta t=64$~ms.  This test resulted in a value of $t_1 =
-0.565$, which corresponds to rejection of the null hypothesis of
independence at a 43\% probability.  Therefore, the test is consistent
with no correlation, although it may also be consistent with the weak 
anticorrelation expected from time dilation (see \S \ref{sec-simul}). 
The results of all of the
tests are summarized graphically in Figure \ref{fig:tablefig}.

\subsubsection{Importance of the Signal-to-Noise Bias}

A correction for the signal-to-noise bias could change these numbers.  Table \ref{table:norris}
shows the results of similar tests performed only on the subset of
bursts for which Dr. J. Norris kindly supplied us with peak-normalized durations.  
This subset consisted of 265 bursts with long ($T \gtrsim 1$~s) durations
and peak count rates greater than 1400~counts~s$^{-1}$, so that the
short duration bias and the problems associated with the
variable threshold rate are minimized.  
As 
expected, tests involving the BATSE $T_{90}$ give much more positive
correlation test results than those involving the 
peak-normalized duration $\hat{T}_{90}$.  The same
trend, but at a much less statistically significant level, 
can be seen in the $\hat{T}_{50}$ results.
It is interesting that this subset of bursts
gives significantly more
negative results than the tests involving the larger sample of bursts
shown in Table \ref{table:fpbar}, no matter which measure of duration
is used.  Since the peak-normalized subsample contains mostly long duration
events, this result could be an indication that long and short duration bursts
have different correlation trends.
In any case, the value of $t_1($data) = -1.61 for the test involving
$T_{eff}(T_{50})$ and $\hat{f}_P(T_{eff})$ for this sample of long duration
bursts is on the verge of being significant,
with the probability of rejecting the null hypothesis at 89\%.
All other numbers indicate less significant rejections, implying
weaker or no anticorrelations.

\subsubsection{Test for Correlation Trends}

It is possible
that the less significant results for the larger sample 
could arise from correlations of
opposite trends in different portions of the data.  For example
equal and opposite relations between $\hat{f}_P$ and say short
($T < 1$~s) and long ($T>1$~s) bursts can cancel each other
out, giving low $t_1($data) values.  We test this possibility
by dividing the data into subsamples.

Figure \ref{fig:twvsduration} shows the values of $t_1($data)
as a function of cutoff in $T_{eff}(T_{50})$.  Both maximum and minimum
cutoffs for $T_{eff}(T_{50})$ are shown, so that for any subdivision
of the data into duration-limited subsets
one may read off the $t_1($subset) values for those
subsets.  Thus for a given value of $T_{eff}$ the middle (lower)
panels give the value for $t_1($subset) for all $T_{eff}$ less (greater) than
the specified value.  
The dotted line shows $-0.565\sqrt{m/296}$, where $m$ is
the number of points in the subset.  Because of the way $t_1$ is defined
(see Efron \& Petrosian 1992 for details), if the average normalized
rank of each point is the same an increase (or decrease) in the number of points
should lead to an increase (or decrease) in the magnitude of $t_1$ by about
a factor
of $\sqrt{m/M}$, where $M$ is the original number of points.
There are two areas where the data seem to deviate from this form.
For bursts above $T_{eff}(T_{50}) \approx 10$~s the $t_1($subset) values
appear to reach significantly negative values ($t_1 < -1.645$), 
while for bursts below
$T_{eff}(T_{50}) \approx 0.1$~s the $t_1($subset) values approach significantly
positive values ($t_1 > 1.645$).
This finding is consistent with the difference in
$t_1$ values given in Tables \ref{table:fpbar} and \ref{table:norris}.
Such behavior could be an indication that short and
long duration bursts have different correlation properties, which
might be the case if long duration bursts showing the time
dilation were cosmological and short
duration bursts consistent with no time dilation (or even possibly time
contraction) were local.  However, the number of bursts in each of
these subsets is small, so that chance cannot be ruled out as a
reason for these differences.

\subsubsection{Interpretation of the Test Results} \label{sec-simul}

The overall test results may be consistent with no correlation, but they
may also be consistent with a very weak anticorrelation.  Since the
expected time dilation factors are of order 2 or less while the 
range of durations spans many orders of magnitude, the question
becomes ``what value of $t_1$ would we expect 
given such a weak correlation?''  
To answer this question we rely on simulations.
We create a large number of simulated durations and peak fluxes,
chosen such that the univariate distributions of $T$ and $\hat{f}_P$ 
closely resemble
those actually observed (and shown in LP).  For our reference simulations
we chose $\hat{f}_P$ such that its 
differential distribution followed a broken power law with a logarithmic
slope of -2.5 above a flux of 20 photons cm$^{-2}$
s$^{-1}$.  At lower fluxes, the logarithmic slope flattens to -2.0.
Independently, $T$ was chosen to be distributed
as two lognormal peaks at 0.1 and 10 seconds, with a standard deviation
of one order of magnitude.  The selection effects which were 
appropriate for the real data were applied to the simulated data,
and the number of data points in each simulation was chosen to match the
number of untruncated real data points (296).  The statistic $t_1($simdata) 
was calculated for each realization and a distribution of $t_1$ was
formed.  Figure \ref{fig:simul} shows this distribution as the 
solid histogram.  As expected, the distribution is approximately normal
with mean 0 and variance 1.  We then repeated the simulations,
adding a small amount of anticorrelation in a power-law fashion such that
the expected observed time dilation factors for the brightness bins 
analogous to those chosen by Norris et al.
(1994) were 1.3, 1.7, and 2.0.  The distributions of $t_1$ for these
simulations are shown as the long-dashed, dotted, and short-dashed
histograms, respectively (their ordinates have been shifted for
clarity).  The mean values of the distributions are -0.127, -0.67,
-1.21, and -1.51.  Clearly, there is significant overlap of all of the distributions,
so that a test result of $t_1$ between about 0 and -1.6 would be
consistent with any of these at about the 10\% level of confidence.  
The conclusion therefore must be that with the current number of
data points, the test cannot at present distinguish
conclusively between no correlation and the weak correlation
signature expected from cosmological time dilation.  
Assuming a very simple correlation form such that the average normalized
rank of each point is the same, in order for
a $t_1 = 0$ result to rule out an observed factor of 2 dilation to the 90\%
confidence level, one would need $296(1.645/1.51)^2 \approx 350$
bursts in the sample.  For a $t_1 = 0$ result to rule out an observed factor of
1.3 dilation to the same level of confidence requires about 1800 bursts!

\subsubsection{Error Correlation Bias} \label{sec-errorbias}

There may exist yet another bias in all of these tests, which 
arises because the errors on quantities such as $\hat{f}_P$ may
be correlated with the values of $T$ due to the dependence
of the $\hat{f}_P$ on $T$.  For example, such a correlation might
increase the spread in $\hat{f}_P$ with decreasing $T$, which
could lead to an edge effect due to the non-symmetric shape of 
the $\hat{f}_P$ distribution.  
We have attempted to estimate the magnitude of this effect on
the measured $t_1$ values through simulations.  We create 
a number of simulated $\hat{f}_P$--$T$ data sets using the method described
in \S \ref{sec-simul} and calculate the
values of $t_1($simdata).  We then randomly shift the data points
by an amount which depends on their particular values of $\hat{f}_P$ and $T$
which is determined by the observational errors in the actual BATSE
data.  Due to this shifting, some points move across the data truncation
boundary into the observable set of data points, while others become
unobservable.  Note that by using this procedure we also take into account
the ``peak flux bias'' caused by Poisson fluctuations in weak bursts.
$t_1($simdata$'$) is recalculated for this error-shifted
data set, and the difference between it and the original statistic is
found.  The process is repeated for a large number of simulated data
sets.  Fortunately, for the specific error properties and data truncations
relevant for the BATSE data it turns out that the largest difference between the
average $t_1$ for the reference simulations and the average $t_1$
for the error-shifted simulations is small ($< 0.1$).
This effect tends to give error-shifted $t_1$ values that are slightly larger than
the true $t_1$ values.  Note that any correction for this effect and for
the correlation between duration and spectral index mentioned previously
in \S \ref{sec-effdur} would tend to reduce the significance of 
any time dilation signature.  Given that the significance of the
test results is  marginal to begin with,
we do not consider correcting the results for these additional effects.

\subsection{Indirect Tests of Correlation} \label{sec-indirect}

It is possible to use the complete three dimensional data set $\hat{f}_P$,
$\hat{f}_{lim}$, and $T_x$ to determine the correlation between
any of the three variables.  The threshold flux $\bar{f}_{lim}$
set prior to the occurrence of a burst is expected to be uncorrelated
with any of the burst properties, in particular $\bar{f}_P$, $\hat{f}_P$,
or $T_x$.  As we shall see below this does not seem to be true for all
cases.  For the moment assuming the expected absence of correlation,
an analysis of the three dimensional data effectively gives the correlation
between $\hat{f}_P$ and $T_x$.  The description of such an analysis
is complicated but it turns out to be unnecessary because it can
easily be reduced to the simpler two dimensional case.  This is 
accomplished by the transformation of $\bar{f}_{lim}$ and $T_x$ into
$\hat{f}_{lim}$ as described by equation (\ref{eq:dilfpflim}).
With this transformation we now have a two dimensional distribution
of $\hat{f}_P$ and $\hat{f}_{lim}$ with the simple truncation
$\hat{f}_P \ge \hat{f}_{lim}$.
Therefore, a test of the correlation between $\hat{f}_P$ and $\hat{f}_{lim}$
can be made without the exclusion of some of the data that was necessary to
produce Tables \ref{table:fpbar} and \ref{table:norris}.  
If the $\hat{f}_P$--$\hat{f}_{lim}$ test gives a strong correlation,
it can be inferred that there is exists an anticorrelation between
$\hat{f}_P$ and $T$, because of the dependence of 
$\hat{f}_{lim}$ on $T$ in equation (\ref{eq:fptrue}).  

Similarly, instead of defining a variable threshold for $\hat{f}_P$ we
may define one for the duration $T_x$.  It can be seen that if we define
(see LP for more details)
\begin{equation}
T_{x,lim}(\hat{f}_P) = \frac{\eta \Delta t}{\hat{f}_P/\bar{f}_{lim} -1}
\end{equation}
then we have a two dimensional data set $T_x$ and $T_{x,lim}$ with
the truncation $T_x \geq T_{x,lim}$.  Because of the dependence of
$T_{x,lim}$ on $\hat{f}_P$, the test of correlation between
$T_x$ and $T_{x,lim}$ will amount to a test of anticorrelation between
$T_x$ and $\hat{f}_P$.

\subsubsection{Peak Flux vs. Duration}

Table \ref{table:full} shows the results of several further
tests utilizing $\hat{f}_P$ and $\hat{f}_{lim}$ or $T_x$ and $T_{x,lim}$.
The results of these tests are generally consistent with the $\hat{f}_P$--$T$
tests.  The only seemingly significant correlations between
measures involving peak flux appear in the 64~ms column.
If taken at face value, 
the fairly large positive values of $t_1$ for $\hat{f}_P$ versus $\hat{f}_{lim}$
would imply a significant dilation effect.  However, this conclusion is
suspect
because $t_1$ for $\bar{f}_P$ versus $\bar{f}_{lim}$
gives a very significant correlation.
This result is puzzling, because as stated above there should not be any
correlation between the limiting flux set prior to the occurrence of a burst
and the 
subsequent peak flux of the triggered burst.
This correlation is unlikely to be caused by a
few outliers, because it persists even when we divide
the bursts up into subsets, either chronologically or in a random
fashion.  

There is the possibility that
for some reason the transformation from counts $\bar{C}_P$ to flux $\bar{f}_P$
is biased in
some systematic way.  
The relationship between the two quantities is 
\begin{equation}
\bar{f}_P = \bar{C}_P/A_{eff}(\theta,\phi), \label {eq:dilarea}
\end{equation}
where $A_{eff}(\theta,\phi)$ is the effective 
area of the detector for the
direction $\theta,\phi$ of the burst, which depends mildly on its spectrum.
In effect, this would be saying that the
effective detector observing area or burst spectrum varies systematically 
with peak count rate or limiting count rate to produce the discrepant
correlation.  
To test this
hypothesis we show in Figure \ref{fig:fpflimbias}
plots of $A_{eff}(\theta,\phi)$ versus average peak count rate, with
the bursts divided up by their limiting count rate.  Due to the
way the detector software operates, most of the bursts are
triggered at one of three discrete values in $\bar{C}_{lim}$.
It can be seen that only bursts with $C_{min} = \bar{C}_{lim}\Delta t = 60$~counts
display a highly significant correlation.  Currently
we have no explanation for this result other than the 
generally unsatisfying explanation that it could be a statistical fluctuation.
This behavior
is not evidenced in $\Delta t=1024$~ms data, and only very marginal evidence
for it is present for $\Delta t=256$~ms data.  Table \ref{table:climbins}
summarizes the results.

It should be noted that the first five values in the last column of
Table \ref{table:full} give essentially the same values of $t_1$.
This result suggests that if the likely spurious effect (which could
be due to an unfortunate and improbable fluctuation) were
eliminated, then there will be little significant correlation
remaining between $\hat{f}_P$ and any of the durations.  Using
the other trigger durations as a guide, it can be seen that 
when the anomalous correlation is absent the $t_1$ values show
no strong evidence for correlation.

\subsubsection{Fluence vs. Duration} \label{sec-flujust}

We now consider the correlation between fluence ${\cal F}$ and duration $T$.
As described in \S \ref{sec-fluencesection}, from the BATSE data 
we can obtain ${\cal F}$, ${\cal F}_{lim} = {\cal F}{C}_{min}/{C}_{max}$,
and some measure of duration, with ${\cal F} \geq {\cal F}_{lim}$.
However, the truncation in the ${\cal F}$--$T$ plane cannot be
obtained directly from the data, so that we cannot directly test
the correlation between these quantities.  However, we do know that
the burst selection process indicates that the threshold on the
fluence will depend on the duration, with the exact relation depending
on the pulse shape.  Clearly for simple pulses ${\cal F}_{lim} \propto 
\hat{f}_{lim} T \propto \bar{f}_{lim}(\Delta t + T)$.  
Therefore, we can test the correlation between
${\cal F}$ and $T$ indirectly by considering the correlation between
${\cal F}$ and ${\cal F}_{lim}$.
The final row in Table \ref{table:full} shows the result of the test between
fluence ${\cal F}$ and ${\cal F}_{lim}$, which indicated the presence
of a significant correlation.
Since the fluence limit becomes approximately
proportional to $T$ for long duration events and is not correlated with
$T$ for short duration events,
a positive correlation between fluence and fluence limit would indicate
a positive correlation between fluence and duration, especially for
long duration bursts.

Taken at face value, the large positive test
values indicate a highly significant correlation,
which is in the opposite sense of that expected from the cosmological time
dilation.  It should be cautioned that the interpretation of 
these fluence results involves many difficulties.  For example,
fluence measures are much more sensitive to background subtraction
than any kind of peak flux measure; it would not be hard to imagine
further systematic biases that might affect the test results.
Furthermore, changes in the burst spectrum with time could have a
significant effect on the fluence while not mattering much for the
determination of the peak flux.  The fluence and the peak flux are
also affected slightly differently by redshifting due to the extra
time factor in the fluence, although this difference would not
be enough to account for the positive results we find.

Assuming these issues can be ruled out as the cause for the positive
test results, inspection of Figure \ref{fig:simul} shows that
even the lowest value of $t_1 = +2.28$ is inconsistent with any time dilation
(even the lowest factor of 1.3) and implies a strong correlation
between ${\cal F}$ and $T$.  Our finding that the slope of the 
$\log{N}$-$\log{{\cal F}}$ has as a break in it that is as sharp
or sharper than the break in the $\log{N}$-$\log{\hat{f}_P}$ distribution
(PL) would seem to indicate that the 
distribution of total radiant energy could be narrower than the distribution
of peak luminosity.  On this basis alone the fluence would seem to be the better
``standard candle,'' but it does not show the anticorrelation
between fluence and duration expected for a static population of standard
candles.  One simple possibility is that bursts might not be of cosmological
origin.  However, this result does not rule out cosmological
scenarios because any relation between ${\cal F}$ and $T$ and the
$\log{N}$-$\log{{\cal F}}$ curve can be fitted by invoking an appropriate
evolution of distributions, luminosities, or number density of bursts.
In fact, for a cosmological population of bursts to be consistent with
both the $\log{N}$-$\log{{\cal F}}$ and the time dilation test results
evolution appears to be required.

\section{Conclusion} \label{sec-dilconclusion}

We have searched for time dilation effects in the BATSE 3B data
by defining several measures of burst strength and duration 
and performing a nonparametric correlation test.  Our treatment
differs from previous treatments in that our test can account
for nontrivial data truncations due to observational selection
biases, allowing us to use a larger sample of bursts.
Table \ref{table:percent} shows the percentage of all 1122 bursts in
the BATSE 3B catalog that were able to be used in each test.
These percentages should be compared to the approximately 20\%
used in previous time dilation tests.

The conclusions of this paper are:

\begin{enumerate}
\item We have confirmed that the observational definition of duration
can have a major influence on correlation test results, and suggest 
effective duration (fluence divided by peak energy flux) as
an appropriate measure for use in time dilation tests. 
\item A nonparametric rank statistic test (Efron \& Petrosian 1992) 
was used to overcome data truncation effects resulting from a
short duration bias in the BATSE data.  These tests utilize a
greater number of bursts (up to 46\% of the 1122 triggered bursts)
than previous investigations and extend the test to short durations.
\item Test results for the correlation between 
duration and peak flux are consistent with no
correlation, but are not currently sensitive enough to rule out
the expected weak correlations.  If existing trends continue, 
the 4B catalog may contain enough data to rule out a factor of 2 dilation.
Ruling out a factor of 1.3 dilation would probably require
a factor of six more bursts than currently available.
\item There appears to be slight evidence for different correlation
properties for short and long duration bursts, but this evidence
is not statistically compelling.
\item An indirect test of the correlation between fluence and duration
indicates a positive correlation, which is inconsistent with simple
no-evolution cosmological scenarios if fluence is a better standard
candle than the peak flux.  This point is discussed more completely in PL.
\end{enumerate}

\acknowledgements

We thank J. Norris for providing us with peak-normalized durations
and useful discussions.  This work was supported by funds from NASA
grants NAGW 2290 and NAG-5 2733.

\clearpage

\begin{table*}
\begin{center}
\begin{tabular}{l|ccc}
& \multicolumn{3}{c}{$\bar{f}_P$ versus:} \\
& $T_{90}$ & $T_{50}$ & $T_{eff}(T_{50})$ \\
\tableline
64~ms (296 bursts) & 1.02 & -0.282 &  -0.406  \\
256~ms (314 bursts) & 2.01 & 1.04 &  0.977  \\
1024~ms (382 bursts) & 2.61 & 1.54 &  0.853  \\
\tableline
& \multicolumn{3}{c}{$\hat{f}_P(T_x)$ versus:} \\
& $T_{90}$ & $T_{50}$ & $T_{eff}(T_{50})$ \\
\tableline
64~ms (296 bursts)     & 0.834 & -0.444 &  -0.565  \\
256~ms (314 bursts)    & 1.77  & 0.748 &  0.666  \\
1024~ms (382 bursts)      & 2.02 & 0.829 &  0.356  \\
\end{tabular}
\end{center}
\caption{
Values of $t_1$ for the correlation between peak fluxes and durations for all
available BATSE bursts.
} \label{table:fpbar}
\end{table*}

\begin{table*}
\begin{center}
\begin{tabular}{l|cccccc}
& \multicolumn{6}{c}{$\bar{f}_P$ versus:} \\
      & $T_{90}$ & $T_{50}$ & $T_{eff}(T_{50})$ & $\hat{T}_{90}$ & $\hat{T}_{50}$ & $T_{eff}(\hat{T}_{50})$ \\
\tableline
64~ms (181 bursts) & 0.134    & -1.20   & -1.56  &   -1.22  & -1.20 & -1.56 \\
256~ms (242 bursts) & 1.09    & 0.153   & 0.113  &   -0.728  & -0.266 & 0.110 \\
1024~ms (288 bursts) & 1.42    & 0.325   & -0.142  &   -0.708  & -0.334 & -0.227 \\
\tableline
& \multicolumn{6}{c}{$\hat{f}_P(T_x)$ versus:} \\
      & $T_{90}$ & $T_{50}$ & $T_{eff}(T_{50})$ & $\hat{T}_{90}$ & $\hat{T}_{50}$ & $T_{eff}(\hat{T}_{50})$ \\
\tableline
64~ms (181 bursts) & 0.102    & -1.23   & -1.61  &   -1.28  & -1.28 & -1.60 \\
256~ms (242 bursts) & 1.08    & 0.113   & -0.031  &   -0.703  & -0.316 & -0.033 \\
1024~ms (288 bursts) & 1.21    & 0.010   & -0.287  &   -0.855  & -0.582 & -0.342 \\
\end{tabular}
\end{center}
\caption{
Values of $t_1$ for the correlation between peak fluxes and durations for 
subsamples of bursts with durations measured by Norris.
} \label{table:norris}
\end{table*}

\begin{table*}
\begin{center}
\begin{tabular}{l|ccc}
& 1024 ms & 256 ms & 64 ms \\
& (514 bursts) & (403 bursts) & (417 bursts)\\
\tableline
$\bar{f}_P$ vs $\bar{f}_{lim}$ & 0.812 & 1.74 & 3.01 \\
%$\hat{f}_P(T_{90})$ vs $\hat{f}_{lim}(T_{90})$ & -1.25  & 1.31 & 2.76  \\
$\hat{f}_P(T_{50})$ vs $\hat{f}_{lim}(T_{50})$ & -0.434  & 1.43 & 2.91  \\
%$\hat{f}_P(T_{eff}(T_{90}))$ vs $\hat{f}_{lim}(T_{eff}(T_{90}))$ & -0.219  & 1.83 & 3.05  \\
$\hat{f}_P(T_{eff}(T_{50}))$ vs $\hat{f}_{lim}(T_{eff}(T_{50}))$ & -0.266  & 1.60 & 3.11  \\
%$T_{90}$ vs $T_{90,lim}(\hat{f}_P)$ & -2.67 & -0.459 & -0.257 \\
$T_{50}$ vs $T_{50,lim}(\hat{f}_P)$ & -1.21 & 1.03 & 0.879 \\
%$T_{eff}(T_{90})$ vs $T_{eff,lim}(\hat{f}_P)$ & -0.621 & 1.16 & 0.908 \\
$T_{eff}(T_{50})$ vs $T_{eff,lim}(\hat{f}_P)$ & -1.01 & 0.953 & 0.895 \\
${\cal F}$ vs ${\cal F}_{lim}$ & 4.26 & 3.35 & 2.28 \\
\end{tabular}
\end{center}
\caption{
Values of $t_1$ for various pairs of parameters indirectly testing time dilation.
} \label{table:full}
\end{table*}

%\begin{table*}
%\begin{center}
%\begin{tabular}{l|ccc}
%& 1024 ms & 256 ms & 64 ms \\
%& (336 bursts) & (302 bursts) & (264 bursts)\\
%\tableline
%$\bar{f}_P$ vs $\bar{f}_{lim}$ & 1.08 & 1.86 & 2.95 \\
%$\hat{f}_P(T_{90})$ vs $\hat{f}_{lim}(T_{90})$ & 0.679  & 2.20 & 3.04  \\
%$\hat{f}_P(T_{50})$ vs $\hat{f}_{lim}(T_{50})$ & 1.10 & 1.97 & 2.91  \\
%$\hat{f}_P(T_{eff}(T_{90}))$ vs $\hat{f}_{lim}(T_{eff}(T_{90}))$ & 1.95  & 2.60 & 3.28  \\
%$\hat{f}_P(T_{eff}(T_{50}))$ vs $\hat{f}_{lim}(T_{eff}(T_{50}))$ & 1.20  & 2.08 & 2.87  \\
%$T_{90}$ vs $T_{90,lim}(\hat{f}_P)$ & -0.769 & 0.364 & -0.233 \\
%$T_{50}$ vs $T_{50,lim}(\hat{f}_P)$ & 0.719 & 1.57 & 0.839 \\
%$T_{eff}(T_{90})$ vs $T_{eff,lim}(\hat{f}_P)$ & 1.34 & 2.02 & 1.13 \\
%$T_{eff}(T_{50})$ vs $T_{eff,lim}(\hat{f}_P)$ & 0.862 & 1.89 & 0.982 \\
%${\cal F}$ vs ${\cal F}_{lim}$ & 4.67 & 2.42 & 0.932 \\
%\end{tabular}
%\end{center}
%\caption{
%$t_1$ values calculated using the maximal number of bursts from the sample
%for which Norris has peak normalized durations.  
%} \label{table:full2}
%\end{table*}

\begin{table*}
\begin{center}
\begin{tabular}{lccc}
$\Delta t$ & $C_{min}$ & $t_1$ & number in sample \\
\tableline
64 ms & 60 & 3.87 & 39 \\
64 ms & 66 & 1.38 & 275 \\
64 ms & 71 & 0.737 & 81 \\
256 ms & 121 & 2.24 & 45 \\
256 ms & 132 & -0.243 & 259 \\
256 ms & 143 & 1.44 & 84 \\
1024 ms & 242 & 0.513 & 88 \\
1024 ms & 264 & 0.512 & 299 \\
1024 ms & 286 & -0.266 & 99 \\
\end{tabular}
\end{center}
\caption{
Tests of correlation between $A_{eff}(\theta,\phi)$ and $\bar{C}_P$,
for subsets with different values of $C_{min}=\bar{C}_{lim}\Delta t$.
} \label{table:climbins}
\end{table*}

\begin{table*}
\begin{center}
\begin{tabular}{l|ccc}
& 1024 ms & 256 ms & 64 ms \\
\tableline
%$C_P$ or $\bar{C}_P$ vs $T_x$ & 42\% & 31\% & 32\% \\
$\hat{f}_P$ or $\bar{f}_P$ vs $T_x$ & 34 & 28 & 26 \\
$\hat{f}_P$ vs $\hat{f}_{lim}$ & 46  & 36 & 37  \\
$T_{x}$ vs $T_{lim}$ & 46 & 36 & 37 \\
${\cal F}$ vs ${\cal F}_{lim}$ & 46 & 36 & 37 \\
\end{tabular}
\end{center}
\caption{
Percentages of all 1122 bursts available for each test.
} \label{table:percent}
\end{table*}

\clearpage

\begin{figure}
\caption{
Comparison of the various definitions of duration.  {\em Top panel:}
The ratio of $T_{90}$ to $T_{eff}(T_{50})$ 
as a function of $\bar{C}_P/\bar{C}_{lim}$.
{\em Center panel:} The ratio of $T_{90}$ to $T_{50}$ as a function
of $\bar{C}_P/\bar{C}_{lim}$.  The dotted line marks the value of the ratio 9/5.
{\em Bottom panel:} The ratio of $T_{50}$ to $T_{eff}(T_{50})$
as a function of $\bar{C}_P/\bar{C}_{lim}$. In all three panels, the solid lines indicate
the best power law fits through the points and are meant to give
an indication of the trends.  The $\Delta t=64$~ms trigger timescale
was used for all of these.
}
\label{fig:tcomp64}
\end{figure}

\begin{figure}
\caption{
Correlation test results for several measures of peak flux versus
duration.  The left panel shows the $t_1$ values for the
correlation between $\bar{f}_P$ and $T_{90}$ (solid line), $T_{50}$
(dotted line), and $T_{eff}(T_{50})$ (dashed line) for three
different trigger timescales;  $t_1$ is a test statistic which
measures the deviation of the data from the null hypothesis that 
the variables are uncorrelated (see \S \protect{\ref{sec-analysis}}
for details).
The right panel is the same except with $\hat{f}_P(T_x)$ instead of
$\bar{f}_P$ and gives systematically lower values of $t_1$, showing
the effect of the short duration bias.
The larger values of $t_1$ for $T_{90}$ in both panels show the effect of the
signal-to-noise bias.
}
\label{fig:tablefig}
\end{figure}

\begin{figure}
\caption{
{\em Top panel:}  Scatter plot of $T_{eff}(T_{50})$ versus $\hat{f}_P(T_{eff})$.  
The line shows the truncation boundary due to the duration bias.
{\em Center panel:} The $t_1$ statistic as a function of minimum
$T_{eff}(T_{50})$, where $t_1$ is a measure of the likelihood of a
significant correlation.  The leftmost value of $t_1$ includes all the bursts,
while all subsequent values include only bursts down to some minimum
$T_{eff}(T_{50})$.  The dotted line shows a function proportional to
${\protect \sqrt{m/M}}$, where $m$ is the number of bursts in the subsample and
$M=296$ is the total number of bursts.  {\em Bottom panel:} Same as
in the center panel, except that $t_1$ is plotted as a function of
maximum $T_{eff}(T_{50})$, so that the rightmost value includes all
of the bursts.
}
\label{fig:twvsduration}
\end{figure}

\begin{figure}
\caption{
Distributions of $t_1$, a statistical measure of the
probability that the data are correlated, obtained from simulations.  The solid
histogram shows the reference simulation with no correlation.
The long dashed, dotted, and short dashed histograms show the
$t_1$ distributions for simulations of correlated variables
corresponding to time dilation factors of 1.3, 1.7, and 2.0.
The vertical lines mark the average value of $t_1$ for each set
of 1000 simulations.  The distributions have been shifted vertically
for clarity.
}
\label{fig:simul}
\end{figure}

\begin{figure}
\caption{
Scatter plots of the effective observing area versus average peak
count rate, separated by limiting count rate, for the $\Delta t=64$~ms
sample.  From top to bottom, the plots are for $\bar{C}_{lim}\Delta t=60,
66,$ and 71, respectively.  The lines are the best straight line fits of 
$A_{eff}(\theta,\phi)=m\log{\bar{C}_P}+b$.
}
\label{fig:fpflimbias}
\end{figure}


\begin{thebibliography}{}

\bibitem[Azzam \& Petrosian 1996]{AP96} 
Azzam, W. J. \& Petrosian, V. 1996, \apj, submitted.

\bibitem[Band et al. 1993]{Ban93}
Band, D. et al. 1993, \apj, 413, 281

\bibitem[Band 1994]{Ban93} 
Band, D. L. 1994, \apjl, 432, L23

\bibitem[Efron \& Petrosian 1992]{EP92} 
Efron, B. \& Petrosian, V. 1992, \apj, 399, 345

\bibitem[Fenimore et al. 1995]{F95} 
Fenimore, E. E., In't Zand, J. J. M., Norris, J. P., Bonnell, J. T.,
\& Nemiroff, R. J. 1995, \apjl, 448, L101

\bibitem[Fenimore 1996]{Fen96}
Fenimore, E. E.  1996, proceedings of the 3rd Huntsville
Gamma-Ray Burst Symposium, in press

\bibitem[Goodman 1986]{Goo86} 
Goodman, J. 1986, \apjl, 308, L47

\bibitem[Kouveliotou et al. 1993]{Kou93} 
Kouveliotou, C., Meegan, C. A., Fishman, G. J., Bhat, N. P., Briggs,
M. S., Koshut, T. M., Paciesas, W. S., \& Pendleton, G. N. 1993, 
\apjl, 413, L101.

\bibitem[LP]{LP96} 
Lee, T. T. \& Petrosian, V. 1996, \apj, 470, XXX (LP)

%\bibitem[Lee et al. 1993]{LPM93} 
%Lee, T. T., Petrosian, V., \& McTiernan, J. M. 1993, \apj, 412, 401

%\bibitem[Lee et al. 1995]{LPM95} 
%Lee, T. T., Petrosian, V., \& McTiernan, J. M. 1995, \apj, in press

\bibitem[Mao \& Paczy\'nski. 1992]{MP92} 
Mao, S. \& Paczy\'nski, B. 1992, \apjl, 388, L45

\bibitem[Meegan et al. 1992]{Mee92} 
Meegan, C. A., Fishman, G. J., Wilson, R. B., Paciesas, W. S., Pendleton, G. N., Horack, J. M., Brock, M. N., \& Kouveliotou, C. 1992, Nature, 355, 143

\bibitem[Meszaros \& Rees 1993]{MR93} 
Meszaros, P. \& Rees, M. J. 1993, \apj, 405, 278

\bibitem[Meszaros \& Rees 1994]{MR94} 
Meszaros, P. \& Rees, M. J. 1994, \apjl, 430, L93

\bibitem[Mitrofanov et al. 1994]{Mit94} 
Mitrofanov, I. G., Chernenko, A. M., Pozanenko, A. S., Paciesas,
W. S., Kouveliotou, C., Meegan, C. A., Fishman, G. J., \& Sagdeev,
R. Z. 1994, in AIP Conference Proceedings 307: Gamma Ray Bursts,
eds. G. J. Fishman, J. J. Brainerd, \& K. Hurley, New York: AIP Press,
p. 187.

\bibitem[Mitrofanov 1996]{Mit96}
Mitrofanov, I. G. 1996, Proceedings of the Third Huntsville Symposium on
Gamma-Ray Bursts, in press

\bibitem[N1]{Nor94} 
Norris, J. P., Nemiroff, R. J., Scargle, J. D., Kouveliotou, C., Fishman, G. J., Meegan, C. A., Paciesas, W. S., \& Bonnell, J. T. 1994, \apj, 424, 540 (N1)

\bibitem[N2]{Nor95} 
Norris, J. P., Bonnell, J. T., Nemiroff, R. J., Scargle, J. D., Kouveliotou, C., Paciesas, W. S., Meegan, C. A., \& Fishman, G. J 1995, \apj, 439, 542 (N2)

\bibitem[Norris 1996]{Nor96}
Norris, J. P. 1996, Proceedings of the Third Huntsville Symposium on
Gamma-Ray Bursts, in press

\bibitem[Norris et al. 1996]{Nor96}
Norris, J. P., Nemiroff, R. J., Bonnell, J. T., \& Scargle, J. D.  1996, 
Proceedings of the Third Huntsville Symposium on
Gamma-Ray Bursts, in press

\bibitem[Paczy\'nski 1986]{Pac86} 
Paczy\'nski, B. 1986, \apjl, 308, L43

\bibitem[Paczy\'nski 1992]{Pac92} 
Paczy\'nski, B. 1992, Nature, 255, 521

\bibitem[Petrosian 1993]{Pet93}
Petrosian, V. 1993, \apjl, 402, L33

\bibitem[Petrosian \& Lee 1996]{PL96}
Petrosian, V. \& Lee, T. T., \apjl, 467, XXX (PL)

\bibitem[Petrosian, Lee, \& Azzam 1994]{PLA94} 
Petrosian, V., Lee, T. T., \& Azzam, W. J. 1994, in AIP Conference
Proceedings 307: Gamma Ray Bursts,
eds. G. J. Fishman, J. J. Brainerd, \& K. Hurley, New York: AIP Press,
p. 93 

\bibitem[Piran 1992]{Pir92} 
Piran, T. 1992, \apjl, 389, L45

\bibitem[Wickramasinghe et al. 1993]{Wic93} 
Wickramasinghe, W. A. D. T., Nemiroff, R. J., Norris, J. P., Kouveliotou, C., Fishman, G. J., Meegan, C. A., Wilson, R. B., \& Paciesas, W. S. 1993,
\apjl, 411, L55

\bibitem[Wijers \& Paczy\'nski 1994]{WP94}
Wijers, R. A. M. J. \& Paczy\'nski, B. 1994, \apjl, 437, L107

\bibitem[Yi \& Mao 1994]{YM94}
Yi, I. \& Mao, S. 1994, \prl, 72, 3750

\end{thebibliography}
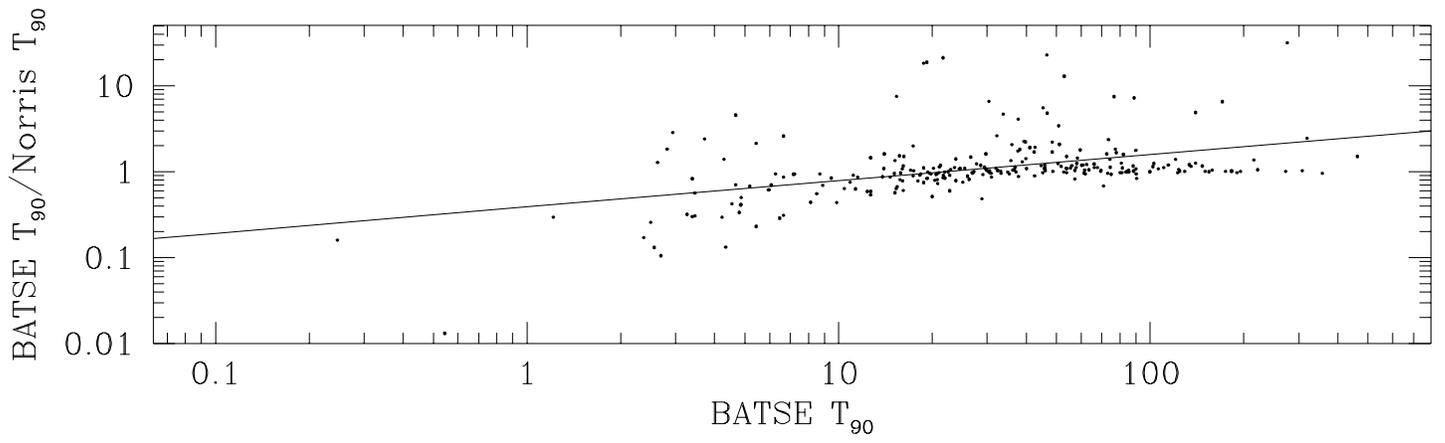
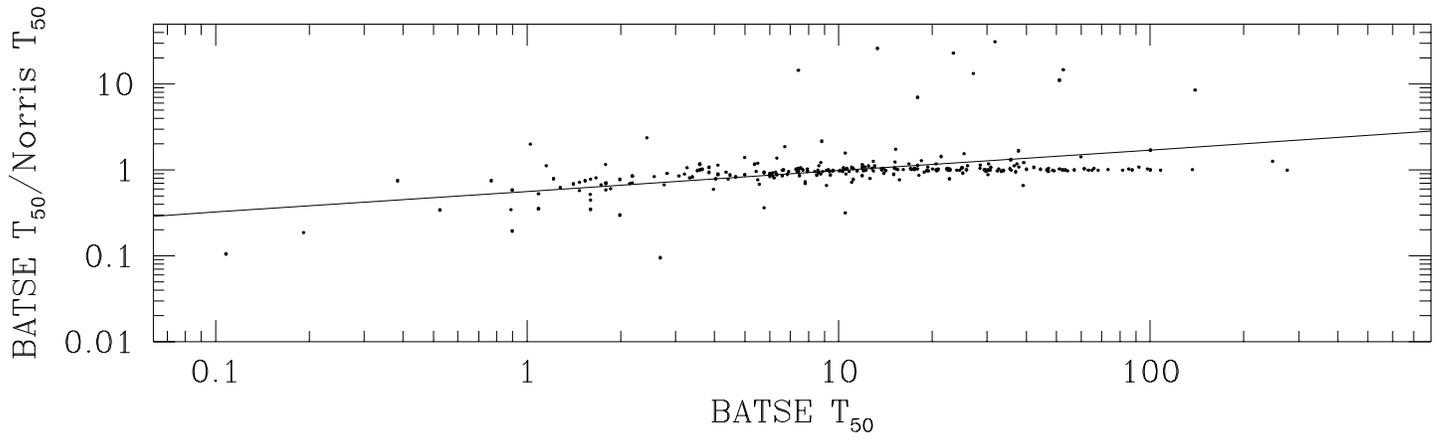
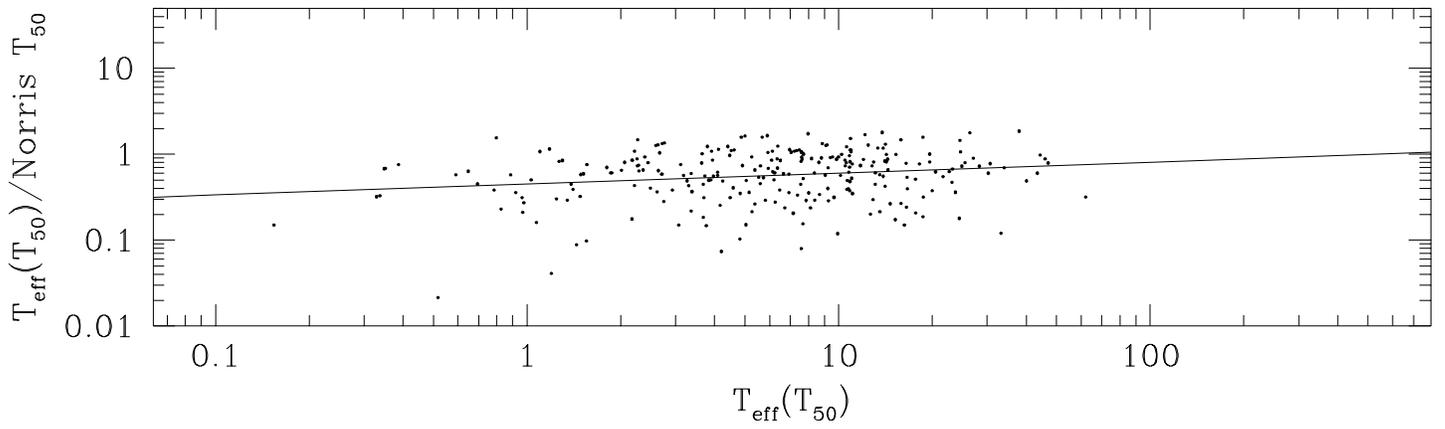
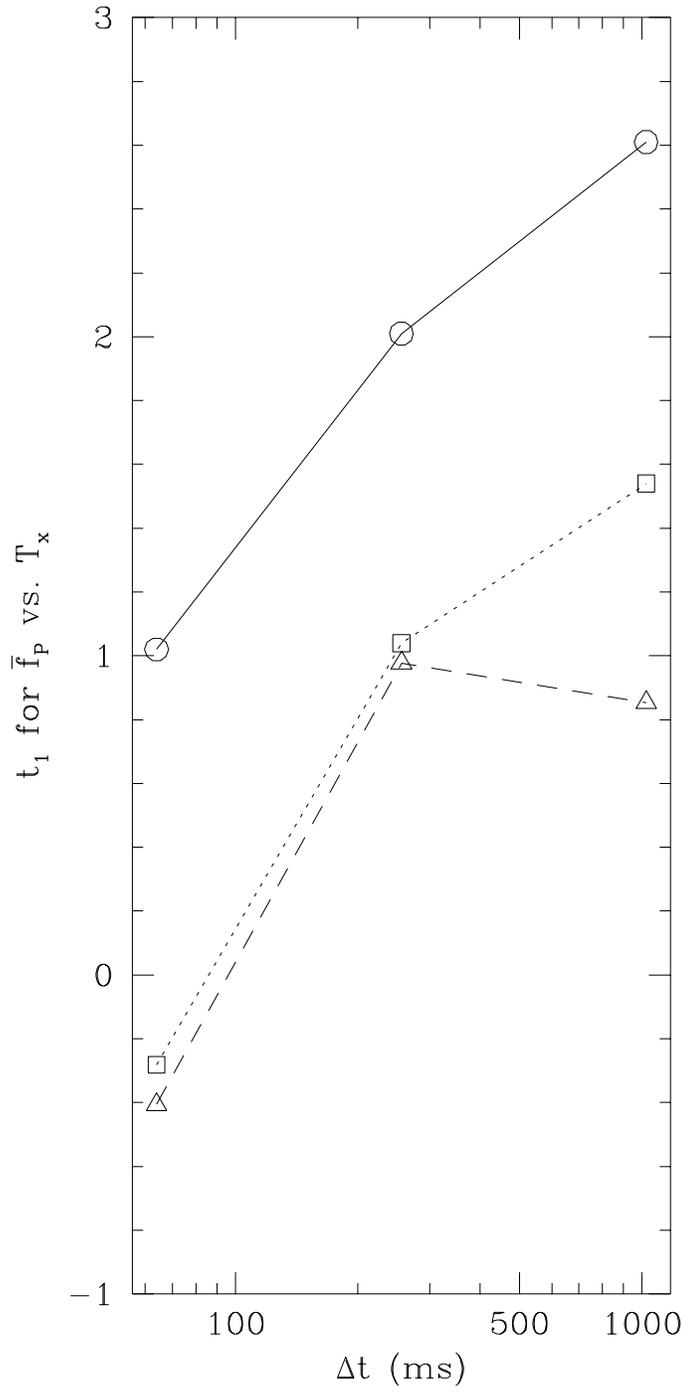
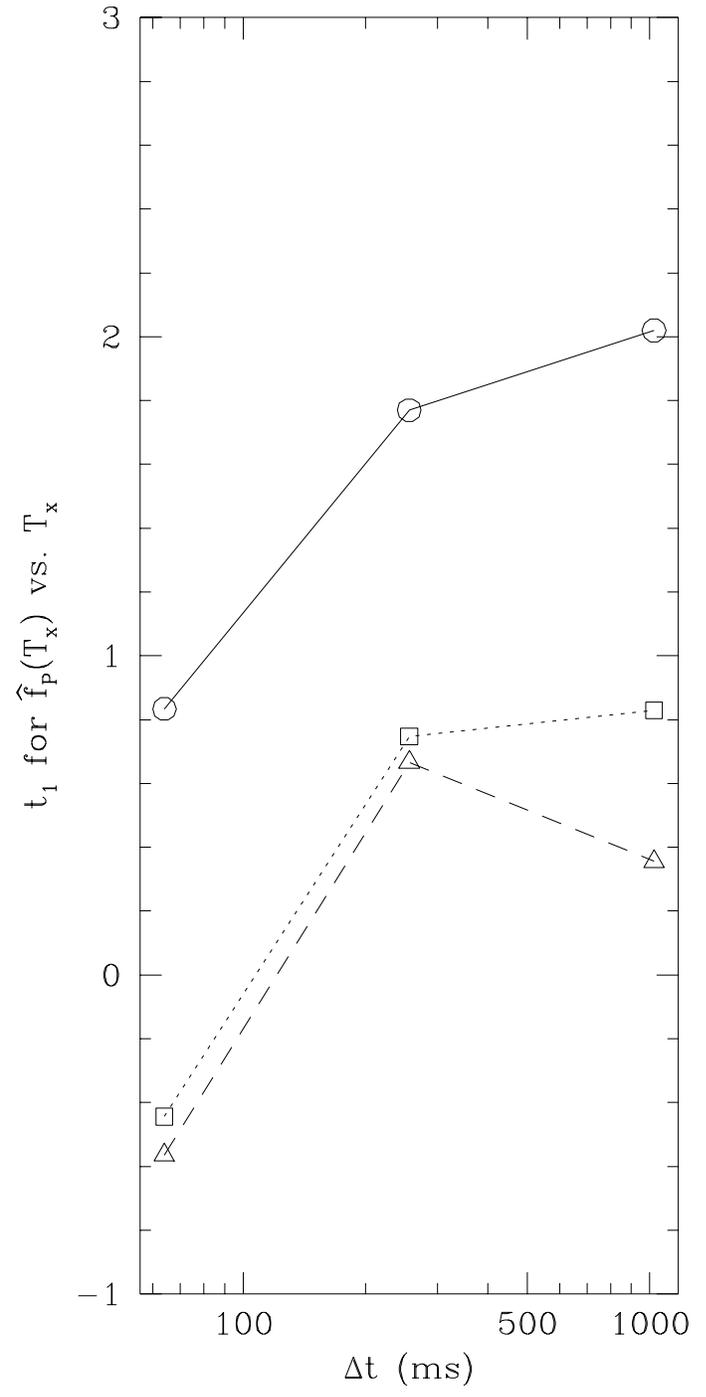
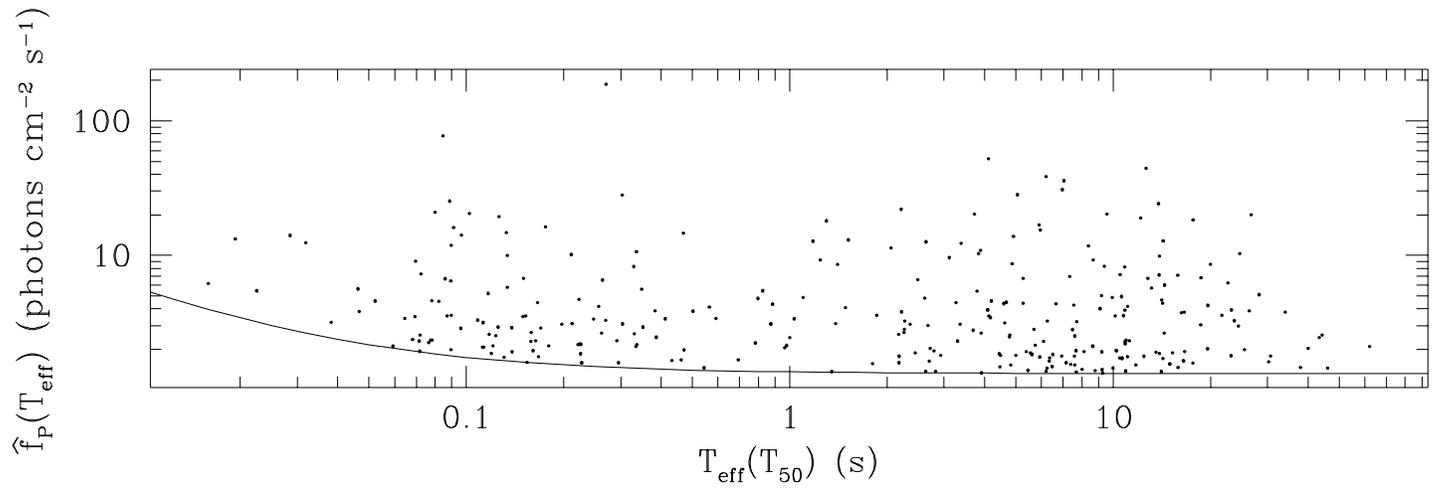
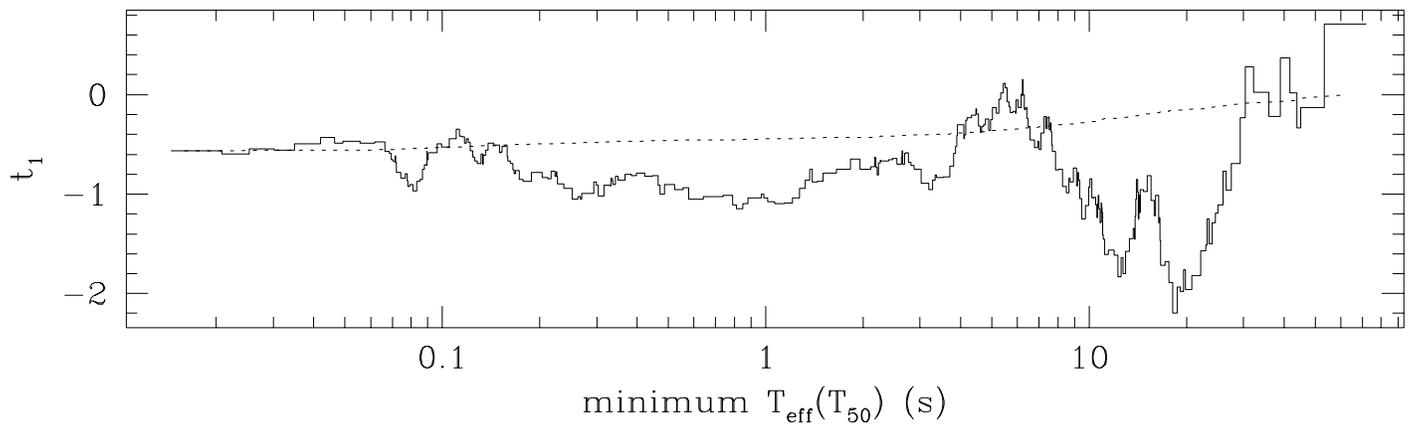
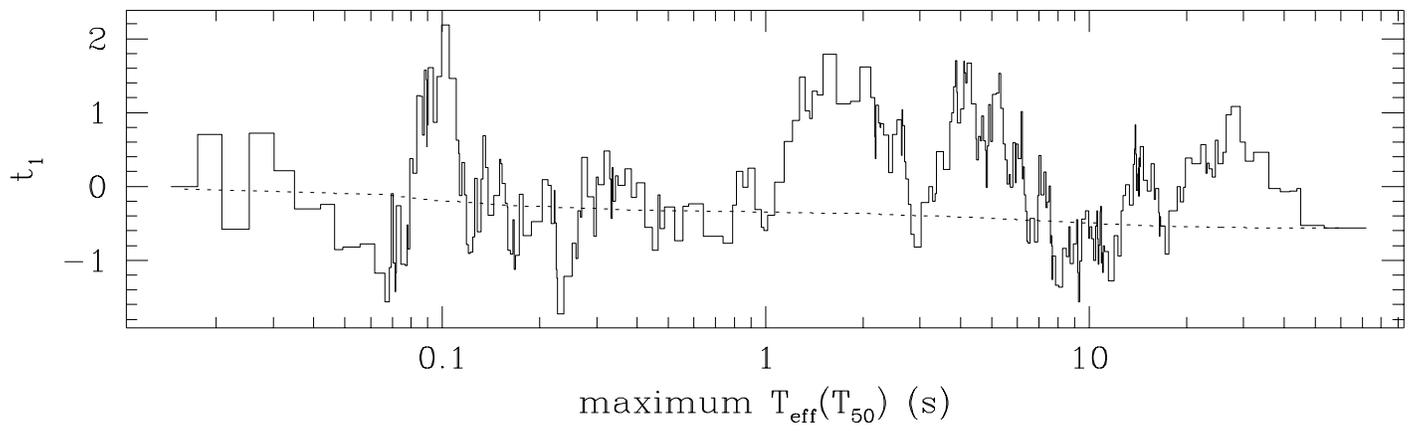
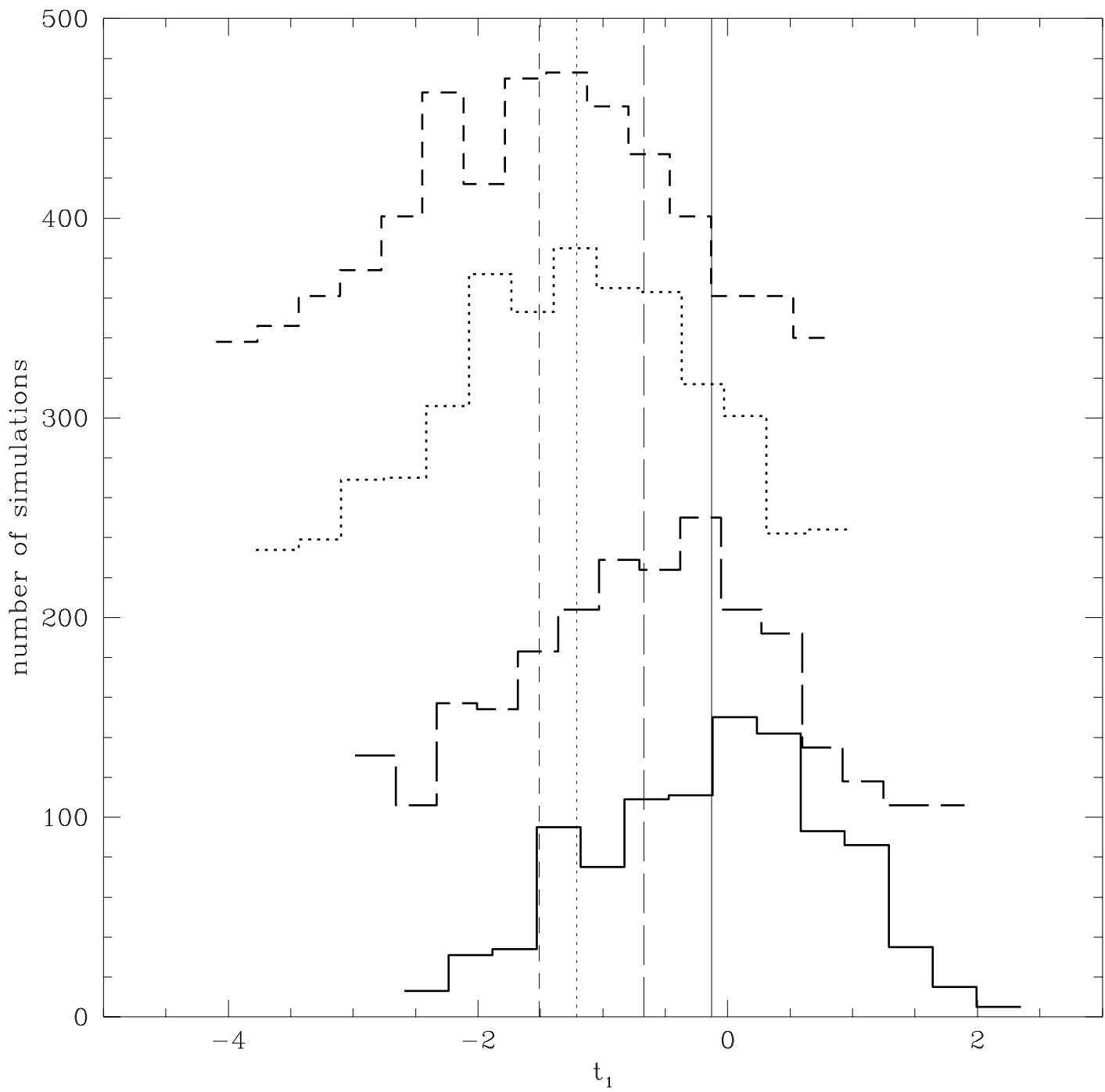
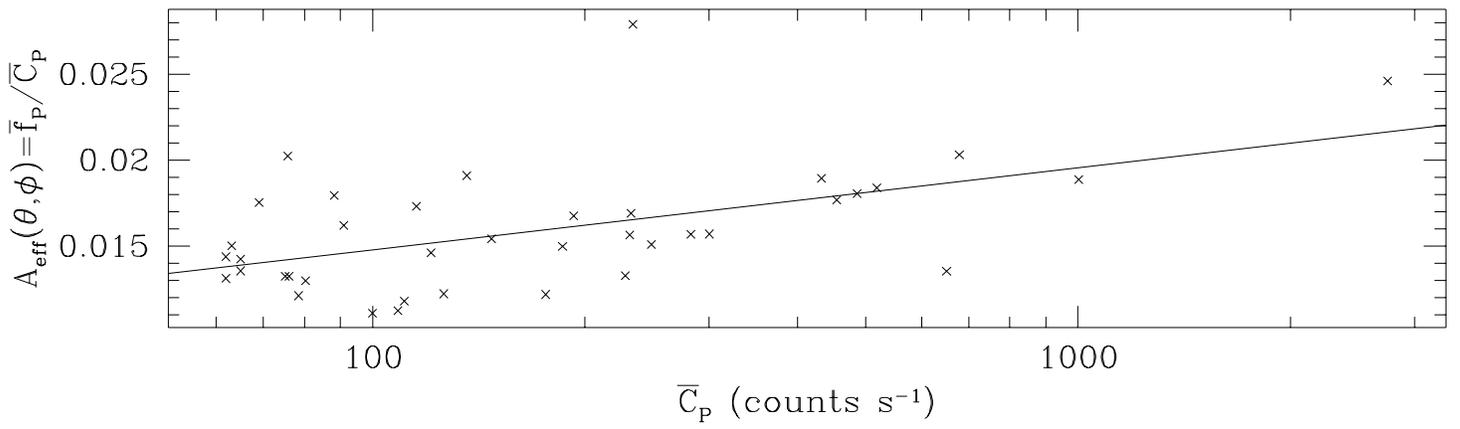
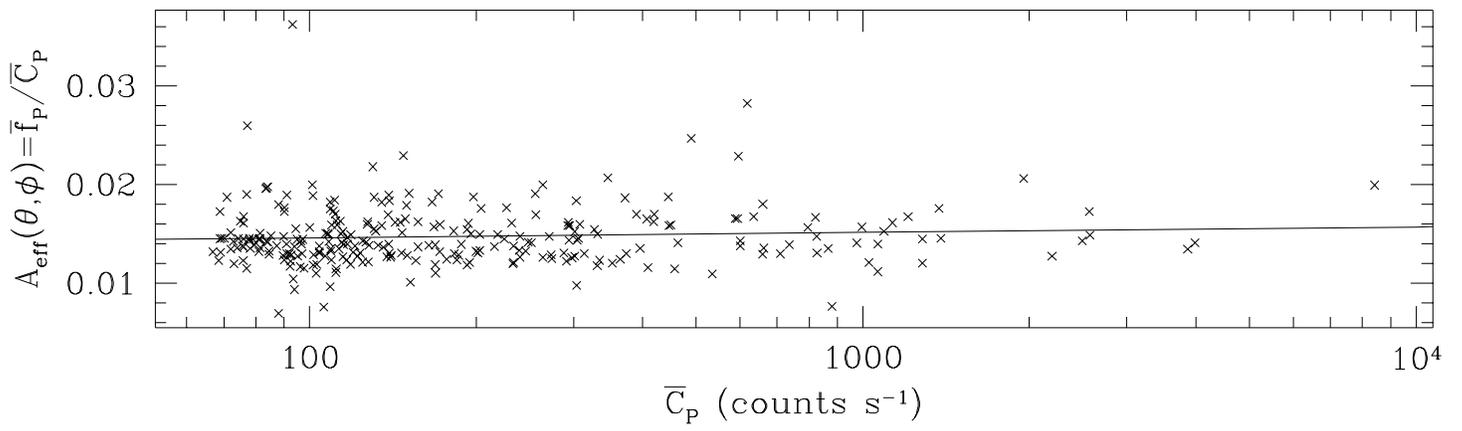
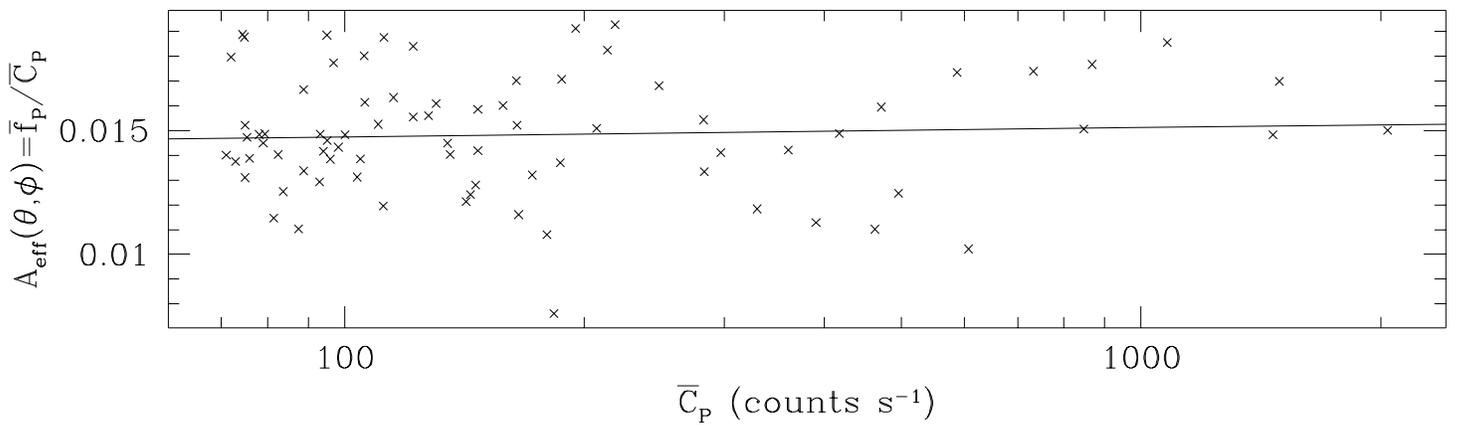
\end{document}